\documentstyle[12pt,aasms4]{article}
\slugcomment{To appear in the Astrophysical Journal as Letter}

\lefthead{Kallenbach et al.}
\righthead{Isotopic Composition of Solar Wind Calcium}

\begin{document}

\title{Isotopic Composition of Solar Wind Calcium: First in situ\\
    Measurement by CELIAS/MTOF on board SOHO}

\author{R. Kallenbach\altaffilmark{1}, F. M. Ipavich\altaffilmark{2}, 
        P. Bochsler\altaffilmark{3}, S. Hefti\altaffilmark{3}, 
        P. Wurz\altaffilmark{3},}
\author{M. R. Aellig\altaffilmark{3}, A. B. Galvin\altaffilmark{2}, 
        J. Geiss\altaffilmark{1}, F. Gliem\altaffilmark{4},
        G. Gloeckler\altaffilmark{2},}
\author{H. Gr\"unwaldt\altaffilmark{5}, M. Hilchenbach\altaffilmark{5}, 
        D. Hovestadt\altaffilmark{6}, and B. Klecker\altaffilmark{6}}

\altaffiltext{1}{International Space Science Institute, Hallerstrasse 6,
                 CH-3012 Bern, Switzerland, reinald.kallenbach@issi.unibe.ch}
\altaffiltext{2}{Dept. of Physics and Astronomy, University of Maryland, 
                 College Park, MD 20742, USA}
\altaffiltext{3}{Physikalisches Institut, University of Bern, CH-3012 Bern, 
                 Switzerland}
\altaffiltext{4}{Institut f\"ur Datenverarbeitungsanlagen, Technische 
                 Universit\"at, D-38106 Braunschweig, Germany}
\altaffiltext{5}{Max-Planck-Institut f\"ur Aeronomie, 
                 D-37189 Katlenburg-Lindau, Germany}
\altaffiltext{6}{Max-Planck-Institut f\"ur Extraterrestrische Physik, 
                 D-85740 Garching, Germany}

\begin{abstract}
We present first results on the Ca isotopic abundances derived from the
high resolution Mass Time-of-Flight (MTOF) spectrometer of the charge,
element, and isotope analysis system (CELIAS) experiment on board the
Solar and Heliospheric Observatory (SOHO). We obtain isotopic ratios  
$^{40}$Ca/$^{42}$Ca $=$ ($128\pm47$) and
$^{40}$Ca/$^{44}$Ca $=$ ($50\pm8$), consistent
with terrestrial values. This is the first in situ determination of the 
solar wind calcium isotopic composition and is important for studies
of stellar modeling and solar system formation since the present-day
solar Ca isotopic abundances are unchanged from their original isotopic
composition in the solar nebula.
\end{abstract}

%\keywords{globular clusters,peanut clusters,bosons,bozos}

\section{Introduction}

The main motivation to study solar wind composition is to obtain information
on the isotopic composition of elements in the Sun. This is important
because the Sun constitutes $99.9\%$ of solar system matter, and for
most elements the solar composition could provide the most reliable 
information on the composition of the primordial solar nebula.  
Because nowhere in the Sun have temperatures ever been high enough to alter 
isotopic abundances of heavy elements by nuclear reactions,
solar Ca is thought to reflect the original isotopic composition in the
solar nebula. Yet, most of what is known of solar isotopic
abundances is inferred from terrestrial and from meteoritic abundances with
a few exceptions: recently, the isotopic composition of solar wind 
magnesium, the first analysis of the isotopic abundances of a refractive
element in solar matter measured in situ with the spacecraft borne mass
spectrometer WIND/MASS, has been reported (\cite{WindMg}).
The solar isotopic composition of the volatile noble gases
helium and neon have been determined in situ with the Apollo foil
experiments (\cite{Geiss}). In situ measurements with
CELIAS/MTOF (\cite{Kallenbach}) on board SOHO have confirmed that the solar
neon isotopic composition differs significantly from the terrestrial
and meteoritic abundances. Isotopic abundances also have been
determined for several elements (C, N, O, He, Ne, Mg) in the higher energy 
(above 10 MeV/amu) solar energetic particles 
(\cite{Leske,Mason,Mewaldt,Selesnick}). From similar measurements 
it has been possible to obtain reliable information on coronal elemental 
abundances including the abundance of calcium (\cite{Breneman}). However, 
the isotopic composition of solar energetic particles (SEP) may not be 
representative of the solar composition because fractionation processes 
occur during the particle acceleration and transport and these processes 
may vary from event to event. The far more fluent solar wind is the
most authentic sample of the solar source composition. From recent 
theoretical models (e.g. \cite{Bodmer}) it is expected that the isotopic 
fractionation in the solar wind flow due to differences in Coulomb drag 
depletes the heavier isotopes by at most a few percent. Therefore 
measurements of the solar wind isotopic abundances of the heavy element
calcium together with previous measurements of magnesium abundances
provide additional evidence that the solar isotopic composition 
of refractive elements agrees with terrestrial and meteoritic values. 

\section{Instrumentation}

\placefigure{fig1}
The MTOF sensor (Figure~\ref{fig1}) of CELIAS (Charge, Element, and
Isotope Analysis System) on board the SOHO (Solar and Heliospheric 
Observatory) spacecraft is an isochronous time-of-flight mass 
spectrometer (\cite{Hovestadt}) with a resolution M/$\Delta$M 
of better than 100. This provides the possibility of resolving the 
different isotopes of almost all solar wind elements in the range from 
$3$ to $60$ atomic mass units (amu). The instrument detects ions at solar 
wind bulk velocities of 300 to 1000 km/s, corresponding to energies 
of about 0.3 to 3 keV/amu. Details on the principle of operation, 
the calibration of the instrument functions, and the format of the data 
transferred by the telemetry can be found in the initial publication on 
CELIAS/MTOF isotope abundance measurements of solar wind 
neon (\cite{Kallenbach}). Here we only briefly describe the instrument 
characteristics which are necessary to understand 
the quantitative evaluation of time-of-flight (TOF) spectra. 

Highly charged solar wind ions enter the instrument (Figure~\ref{fig1}) 
through the WAVE (Wide Angle Variable Energy) entrance system 
that has an energy-per-charge ($E/q$) acceptance bandwidth of about half 
a decade, and a conic field of view of $\pm 25^{o}$ 
width (\cite{Hovestadt}). All calcium isotopes have 
approximately the same bulk velocity, the same charge state, and the 
same width of the drifting maxwellian velocity distribution so that the 
heavier isotopes have a higher center $E/q$. The strongest instrumental 
fractionation occurs when the center $E/q$ of either isotope coincides
with one of the edges of the $E/q$-acceptance of the WAVE. The acceptance
function including the ion optical effects of the postacceleration voltage
to the time-of-flight section has been well
analyzed in the calibration system for mass spectrometers
(\cite{Steinacher,Kallenbach}) of the University of Bern (UoB).
Therefore the flight data can reliably be filtered with respect
to the ion optical instrument discrimination for any element or isotope
detected in the solar wind. A much weaker instrumental fractionation is due 
to the element specific detection efficiencies of the VMASS subsystem which 
is a V-shaped isochronous time-of-flight spectrometer (\cite{Hovestadt}).
The ratio of the Double Coincidence Rate over the Front Secondary 
Electron Detection Assembly Rate has been measured as a function of
ion energy per mass with a $^{40}$Ca$^{4+}$ beam of the MEFISTO
laboratory at UoB (\cite{Marti}). The model functions 
reproduce the calibration data very well so that they can be applied to 
derive the differences in the detection efficiencies of the isotopes
$^{40,42,44}$Ca (Figure~\ref{figca2}). With very good confidence the
uncertainty of these detection efficiencies can be assumed to be lower
than the statistical error evaluated (see Table~\ref{fit}).
\placefigure{figca2}

\section{Data Analysis}

\placefigure{figca3}
Figure~\ref{figca3} shows a spectrum where all pulse height analysis
words of the full year 1996 have been collected in the mass range 39 to
45 amu. TOF channel contents have been put into bins six channels wide
and the uncertainties of the data points have been estimated by the 
linear-optimization algorithm of Rauch-Tung-Striebel (\cite{Gelb}) where 
the filter width corresponds to half the TOF resolution of the instrument;
this filter introduces a minor reduction of the peak resolution of $12\%$. 
This algorithm has been applied to improve the signal-to-noise ratio by a 
factor of four so that $^{42}$Ca$^{+}$ and $^{44}$Ca$^{+}$ can be identified 
with high significance. Note that most incident ions leave the carbon foil
as neutrals or singly charged. The $^{40}$Ca$^{+}$ distribution is 
asymmetrically broadened. It exhibits a different shape than $^{20}$Ne$^{+}$ 
(\cite{Kallenbach}) for one main reason: singly charged calcium has 
generally much higher $E/q$ inside the VMASS than singly charged neon so 
that calcium follows a longer trajectory in the field free region above the 
ion stop microchannel plate (IMCP) detector (\cite{Hovestadt}). Longer 
trajectories in the field-free region correspond to longer times of flight.
This leads to the tails of higher channels in the TOF distribution of all 
the mass peaks in Figure~\ref{figca3}. As in the case of 
$^{20,21,22}$Ne$^{+}$ (\cite{Kallenbach}) the electronic ringing peaks
are visible at about 30 TOF channels to the left of the $^{40,42,44}$Ca$^{+}$
signatures; they arise from reflected start signals. The TOF counts have 
been weighted by the detection efficiencies for the different calcium 
isotopes within the window of $\pm \ 15\%$ instrument fractionation between 
$^{44}$Ca$^{+}$ and $^{40}$Ca$^{+}$. All three detectable calcium peaks, 
$^{40}$Ca$^{+}$, $^{42}$Ca$^{+}$, and $^{44}$Ca$^{+}$, are described by the 
same model function with identical asymmetric shape. Also, the marginally 
detectable or non-detectable peaks $^{39,41}$K$^{+}$ and $^{43}$Ca$^{+}$ are 
included into the fit function, assuming terrestrial isotopic abundances:
\begin{eqnarray}
F(t) = {\sum_{i=39}^{44}}{A_{i}} \times {({exp[-{{(t-p_{i})^{2}}\over{2s^{2}}}]} } & \nonumber\\
       { + {\sum_{j=1}^{5}}{{b_j}\over{1+{{{(t-p_{i}-jl)}^{2}}\over{w^{2}}}}} } & \nonumber\\
       { + {{c}\times{exp[-{{{(t-p_{i}+r)}^{2}}\over{2u^{2}}}]}})}. \nonumber \\
%\eqnum{1}
\label{Eqn}
\end{eqnarray}
$t$ is the time-of-flight channel number and the free fit parameters are
$A_{40,42,44}$ representing the amplitudes of the $^{40,42,44}$Ca$^{+}$
peaks, $p_{40}$ the main Gaussian peak position of $^{40}$Ca$^{+}$, 
$l$ the shift of the Lorentzians describing the asymmetry against the 
main Gaussians for all three peaks of $^{40,42,44}$Ca$^{+}$, 
$r$ the shift of the ringing peaks against the main Gaussians, 
the $b_j$ the relative amplitudes of the shifted Lorentzians compared 
to the main Gaussians, 
$c$ the relative amplitude of the ringing peaks, 
$s$ the width of the main Gaussian peaks, 
$w$ the width of the Lorentzian peaks, and
$u$ the width of the ringing peaks.
The Gaussian peak positions $p_{39,41,42,43,44}$ of $^{39,41}$K$^{+}$ 
and $^{42,43,44}$Ca$^{+}$ are kept fixed at their calibrated TOF channels. 
The total amplitudes $A_{40,42,44}$, the widths $w$ and $u$, 
and the relative amplitudes $b_j$ give the abundances for 
$^{40,42,44}$Ca$^{+}$. This analysis has been done twice assuming that the
solar wind calcium is either tenfold or elevenfold charged.
\placetable{fit}
Table~\ref{fit} lists the integrated counts for 
$^{40,42,44}$Ca$^{+}$ assuming tenfold charged solar wind calcium. This 
gives the count ratios 
$^{40}$Ca$^{+}$/$^{42}$Ca$^{+}$ $=$ ($100\pm25$) and
$^{40}$Ca$^{+}$/$^{44}$Ca$^{+}$ $=$ ($45\pm4$). Analogously, assuming
elevenfold charged solar wind calcium we find 
$^{40}$Ca$^{+}$/$^{42}$Ca$^{+}$ $=$ ($141\pm36$) and
$^{40}$Ca$^{+}$/$^{44}$Ca$^{+}$ $=$ ($53\pm4$).
We further correct these ratios for the effect of heavy elements such
as Si and Fe having different mean velocities in the solar wind
compared to the proton velocity $v_H$ derived from the proton monitor
data. For Ca the mean velocity can be estimated to be 1.027 x $v_H$ - 
14.6 km/s (\cite{Hefti}). This correction in the mean velocity and therefore
in the mean E/q leads to the values 
$^{40}$Ca$^{+}$/$^{42}$Ca$^{+}$ $=$ ($107\pm25$) and
$^{40}$Ca$^{+}$/$^{44}$Ca$^{+}$ $=$ ($46\pm4$) for tenfold charged solar
wind calcium and $^{40}$Ca$^{+}$/$^{42}$Ca$^{+}$ $=$ ($148\pm36$) and
$^{40}$Ca$^{+}$/$^{44}$Ca$^{+}$ $=$ ($54\pm4$) for elevenfold charged 
solar wind calcium. According to~\cite{Aellig} and~\cite{Kern} the coronal 
freeze-in temperatures for the time period considered in this paper are 
such that the average solar wind calcium charge state should be somewhere 
between ten and eleven; the average number is approximately 10.5.
Therefore the final values for the solar wind calcium isotopic abundance 
ratios are taken as the average of the two values derived for the two 
assumptions of tenfold and elevenfold calcium. The error of the mean value 
is chosen sufficiently large to be compatible with both assumptions. 
For all ratios an additional uncertainty of $4\%$ is due to uncertainties
in the absolute calibration of the instrument acceptance. This results in
$^{40}$Ca$^{+}$/$^{42}$Ca$^{+}$ $=$ ($128\pm47$) and
$^{40}$Ca$^{+}$/$^{44}$Ca$^{+}$ $=$ ($50\pm8$). More precise values can
be determined, once the charge distribution of solar wind calcium is better
known for the year 1996 with other instrumentation on board the SOHO 
spacecraft. The sensors WIND/MASS and CELIAS/CTOF will provide sufficient 
information to reevaluate the calcium data.

\section{Results}

Based on the assumption that the solar wind calcium charge state varies 
between ten and eleven as a long term average, we find that the isotopic 
abundance ratios of solar wind calcium are 
$^{40}$Ca/$^{42}$Ca $=$ ($128\pm47$) and
$^{40}$Ca/$^{44}$Ca $=$ ($50\pm8$). Both ratios are consistent with the
terrestrial ratios $^{40}$Ca/$^{42}$Ca $=$ ($151.04\pm0.02$) and
$^{40}$Ca/$^{44}$Ca $=$ ($47.153\pm0.003$) (\cite{Russell}). More 
precise values can be obtained once a time series during the year 1996 
of the calcium charge states in the solar wind at the SOHO location is 
available.

\section{Discussion}

From astrophysical and geochemical considerations it can be concluded that
the isotopic composition of photospheric Ca must be within small fractions
of per mills identical to the terrestrial composition. The similarity
of the solar wind results with the terrestrial values suggests the 
preliminary conclusion that isotopic fractionation within the solar
wind plays a minor role. This is supported by the model calculations 
of~\cite{Bodmer} that consider differences in the Coulomb drag
for isotopes of various elements. Fractionation effects not larger than a
few percent are expected for Ne and Mg. For the heavy element Ca even weaker
isotopic fractionation has to be expected. Although there are as yet no
detailed evaluations of a more extensive time series with calcium isotopic 
ratios to be determined in different solar wind regimes, we suggest that 
the Ca abundance discussed in this work is within a few percent of the true 
solar composition.

Table~\ref{comparison} shows a comparison between the terrestrial,
solar and presolar grain calcium isotopic composition. There is no
evidence for large variations in these isotopic compositions except
for the case of the presolar X-grains that
originate from supernovae explosions (\cite{Hoppe,Nittler}). The large
enrichment in $^{44}$Ca in the X-grains is a consequence of the
radioactive decay of $^{44}$Ti in supernovae material.
\placetable{comparison}

\acknowledgments

This work was supported by the Swiss National Science Foundation, by the
PRODEX program of ESA, by NASA grant NAG5-2754, and by DARA, Germany, with
grants 50 OC 89056 and 50 OC 9605. The flight-spare unit of MTOF has been
recalibrated with the electron cyclotron resonance ion source in the 
MEFISTO laboratory at the University of Bern and with
the support of Adrian Marti and Reto Schletti.

\clearpage

\begin{deluxetable}{crrrrrrrrrrr}
\footnotesize
\tablecaption{Fitted parameters of the model function (Equation~\ref{Eqn}) 
 for the calcium PHA spectrum. \label{fit}}
\tablewidth{0pt}
\tablehead{
\colhead{} &
\colhead{$A_{i} \ (10+)$\tablenotemark{a}} &
\colhead{$A_{i} \ (11+)$\tablenotemark{b}} &
\colhead{$p_{i}$\tablenotemark{c}} &
\colhead{Total Counts (10+)\tablenotemark{a}} &
\colhead{Total Counts (11+)\tablenotemark{b}}
}
\startdata
$^{40}$Ca$^{+}$ $ \ $ &$9756\pm42$ &$9600\pm41$ &$3053.1$ &$58289\pm251$ &$57357\pm245$\nl
$^{42}$Ca$^{+}$ $ \ $ &$98\pm25$ &$68\pm17$ &$3128.4$ &$585\pm149$ &$406\pm102$\nl
$^{43}$Ca$^{+}$ $ \ $ &$14\pm14$ &$11\pm11$ &$3165.5$ &$84\pm84$ &$66\pm66$\nl
$^{44}$Ca$^{+}$ $ \ $ &$218\pm18$ &$183\pm15$ &$3203.9$ &$1302\pm108$ &$1093\pm90$\nl
$^{39}$K$^{+}$ $ \ $ &$148\pm14$ &$103\pm12$ &$3014.6$ &$884\pm84$ &$615\pm72$\nl
$^{41}$K$^{+}$ $ \ $ &$11\pm11$ &$7\pm7$ &$3091.0$ &$64\pm64$ &$45\pm45$\nl
 
\enddata
\tablenotetext{a}{Data reduction based on the assumption of tenfold charged
solar wind calcium.}
\tablenotetext{b}{Data reduction based on the assumption of elevenfold
charged solar wind calcium.}
\tablenotetext{c}{Fixed calibrated relative peak positions with fitted
offset to match spectrum of flight data.}
\tablecomments{The $1\sigma$-errors of the amplitudes are 
determined with the method of maximum likelihood. The parameters that are
equal for all isotopes and for both assumptions of ten- and elevenfold
charged solar wind calcium have been fitted as $s$ $=$ $9.222$,
$b_1$ $=$ $0.023$, $b_2$ $=$ $0.009$, $b_3$ $=$ $0.047$, $b_4$ $=$ $0.178$,
$b_5$ $=$ $0.117$, $l$ $=$ $9.454$, and $w$ $=$ $10.146$. The values for
the ringing peaks have been taken from the neon evaluations and kept
fixed: $c$ $=$ $0.035$, $r$ $=$ $31.66$, and $u$ $=$ $5.71$.}

\end{deluxetable}

\clearpage

\begin{deluxetable}{crrrrrrrrrrr}
\footnotesize
\tablecaption{Results from measurements of the terrestrial, solar,
and presolar calcium isotopic composition \label{comparison}}
\tablewidth{0pt}
\tablehead{
\colhead{$^{40}$Ca/$^{44}$Ca} &
\colhead{$^{40}$Ca/$^{42}$Ca} &
\colhead{Source} &
\colhead{Reference}
}
\startdata
$47.153\pm0.003$ &$151.04\pm0.02$ &terrestrial &\cite{Russell}\nl
$45.1\pm0.8$ &$147\pm2$ &average of presolar grains &\cite{Amari}\nl
$2.3\pm0.2$ &... & presolar X-grain &\cite{Hoppe}\nl
$50\pm8$ &$128\pm47$ &solar wind (SOHO/CELIAS) &this work\nl
 
\enddata

\end{deluxetable}

\clearpage

\figcaption[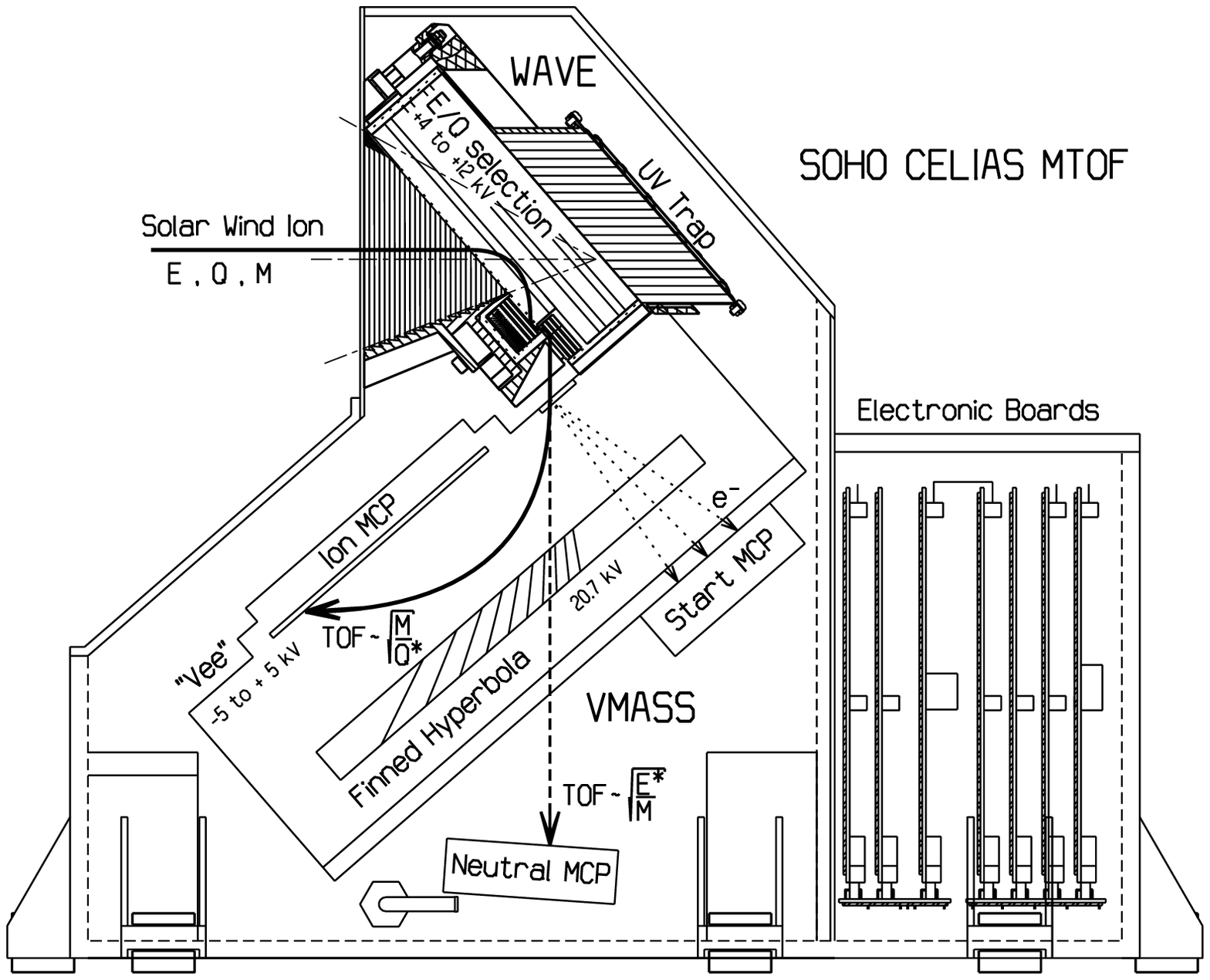]{Schematic view of the
 SOHO/CELIAS/MTOF sensor. \label{fig1}}

\figcaption[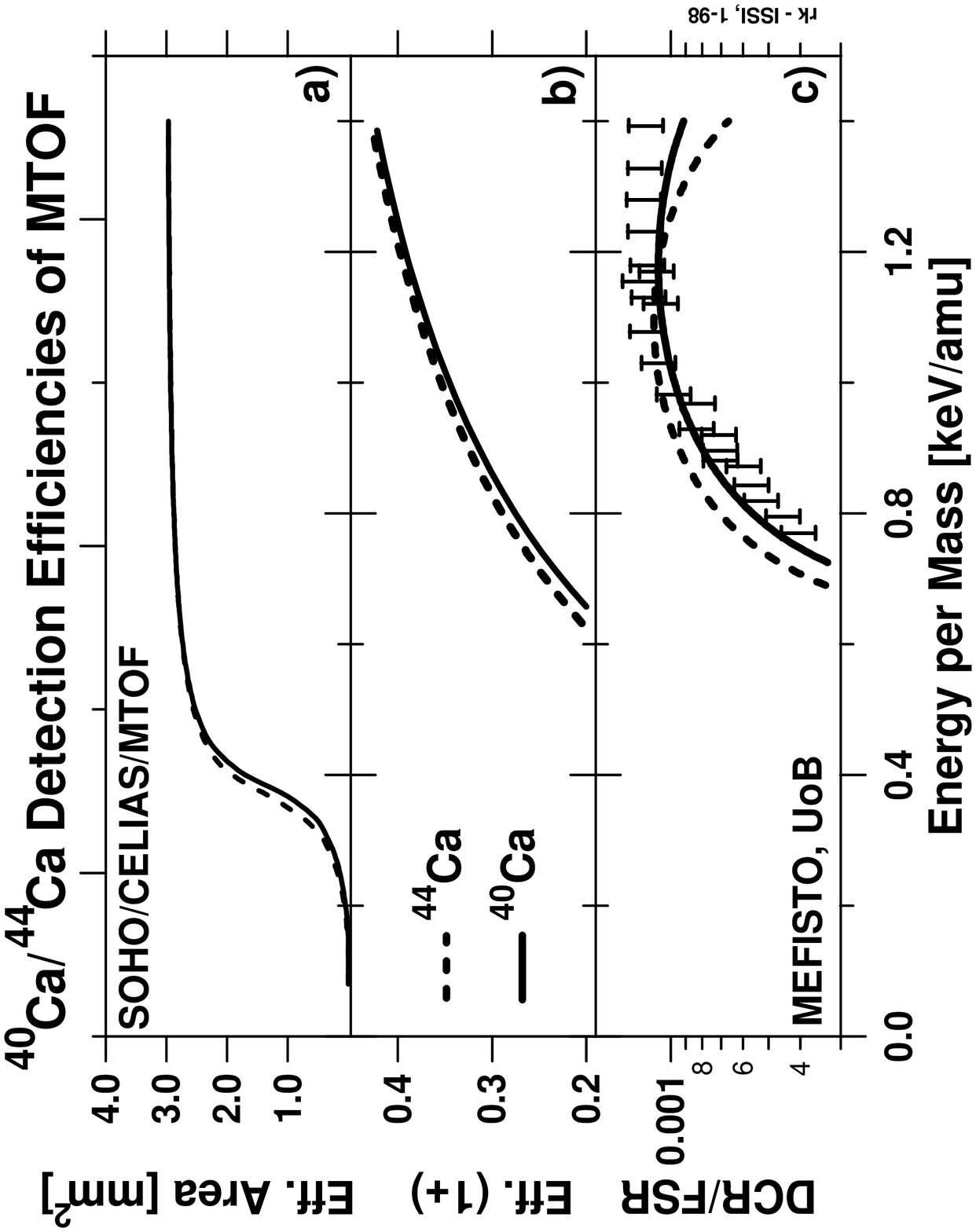]{a) Front Secondary Electron Detection Assembly
Rate (FSR) model functions versus energy per mass in units of mm$^{2}$ 
representing the effective detection area of the CELIAS/MTOF sensor at 
optimum orientation and energy-per-charge acceptance. b) Illustrates the 
probability that the $^{40}$Ca and $^{44}$Ca isotopes leave the foil singly 
ionized. Most particles leave the carbon foil at the entrance of the 
isochronous time-of-flight spectrometer VMASS as neutrals or singly charged. 
c) Double Coincidence Rate (DCR) of $^{40}$Ca$^{+}$ and $^{44}$Ca$^{+}$
over FSR versus ion energy-per-mass ratio:
the calibration data from the MEFISTO ion source at UoB (\cite{Marti}) are
described by a model that includes charge exchange processes, energy loss 
and straggling, including its corresponding angular scattering in the 
carbon foil, ion optics, and the stop detector efficiency. At the low
energies $^{44}$Ca$^{+}$ has a higher chance to be detected than
$^{40}$Ca$^{+}$ because the heavier isotope suffers less angular straggling,
whereas at the higher energies $^{44}$Ca$^{+}$ has a higher risk to overshoot
the ion stop micro channel plate (IMCP) or to hit the hyperbola deflection 
electrode than $^{40}$Ca$^{+}$ because the heavier particle suffers less 
energy loss.\label{figca2}}

\figcaption[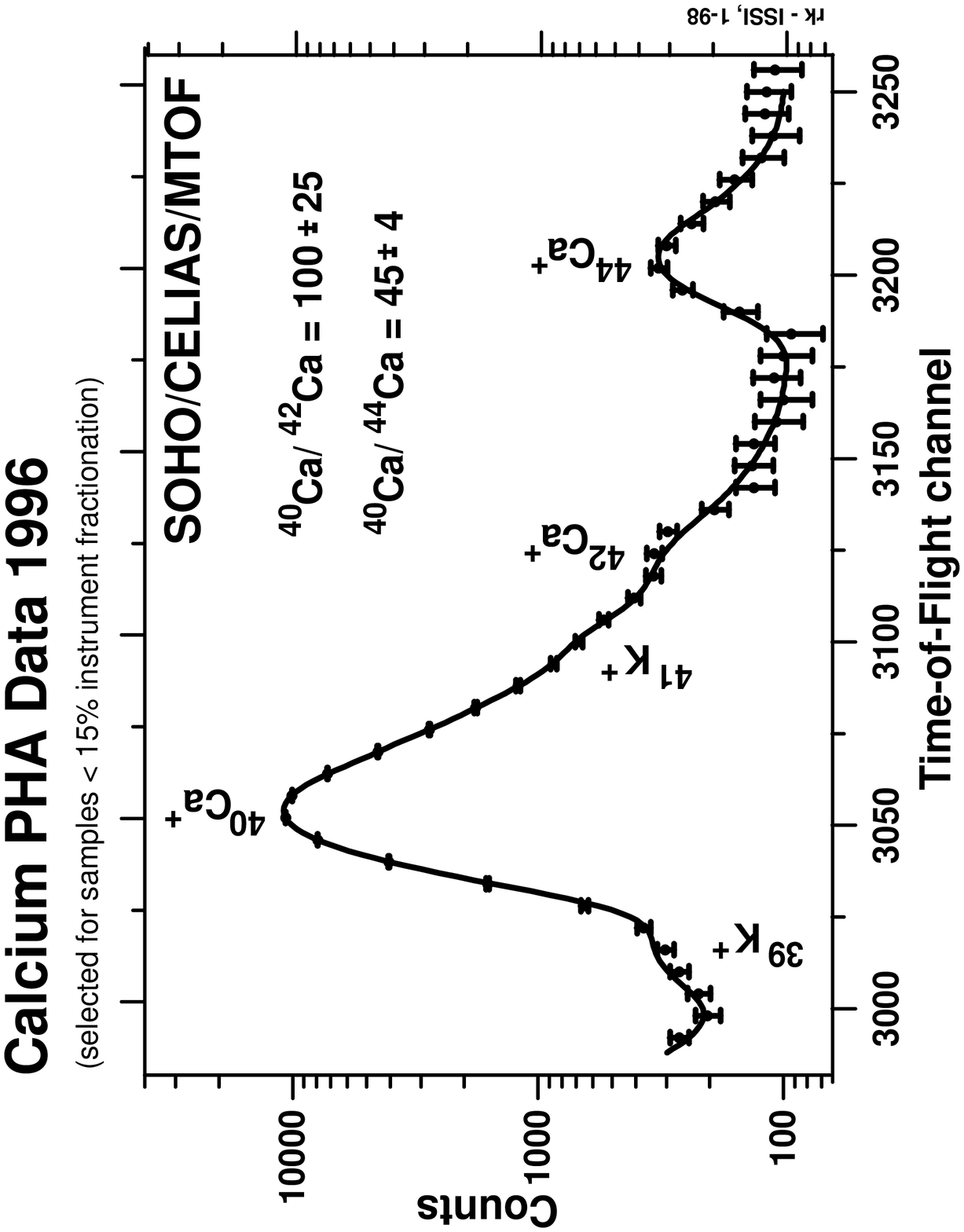]{
Time-of-flight spectrum derived from the pulse height analysis words of 
the full year 1996 filtered with respect to instrument fractionation with 
a window of $\pm$15$\%$. A background of $1500$ counts has been subtracted 
in the diagram but has been included in the error estimation with the 
linear-optimization algorithm of Rauch-Tung-Striebel (\cite{Gelb}). 
The solid line represents the fitted model function of the TOF response. 
The analysis shown here assumes that the solar wind calcium always entered 
the instrument tenfold charged. The same analysis has been redone assuming 
elevenfold charged solar wind calcium (see text).\label{figca3}}
\clearpage
\plotone{figpap1.eps}
\clearpage
\plotone{figca2.eps}
\clearpage
\plotone{figca3.eps}
\end{document}